\documentclass[]{jfm}
\usepackage{graphicx}
\usepackage{epstopdf, epsfig}
\usepackage{multirow}
\usepackage{mathrsfs,amsmath}
\usepackage{bm}
\usepackage{amssymb}
\usepackage{multirow,bigdelim}
\usepackage{upgreek}
\usepackage{units}
\usepackage{array,multirow}
\usepackage{booktabs}
\usepackage{psfrag}
\usepackage{tikz}
\usetikzlibrary{arrows.meta}
\usepackage{color}
\usepackage{xcolor}
\usepackage{threeparttable}	
\usepackage{dcolumn}		
\newcolumntype{d}[1]{D{.}{.}{#1}}
\usetikzlibrary{decorations}
\usetikzlibrary{decorations.markings}
\usetikzlibrary{calc,decorations.pathreplacing}
\usepackage{lipsum}
\usepackage{mathtools}
\usepackage{xfrac}
\usepackage{ar}
\usepackage{enumitem}
\usepackage{url}
\usepackage{subcaption}
\usepackage{mwe}
\usepackage{float}
\usepackage{lmodern}
\usepackage{soul}

\newcommand{\rev}[1]{\textcolor{black}{#1}}

\shorttitle{The large-scale footprint in small-scale Rayleigh-B\'enard turbulence}
\shortauthor{P. Berghout, W.J. Baars and D. Krug}

\title{The large-scale footprint  in small-scale Rayleigh-B\'enard turbulence}

\author{Pieter Berghout\aff{1}, 
  Woutijn J. Baars\aff{2}
 \and Dominik Krug\aff{1}
  \corresp{\email{d.j.krug@utwente.nl}}}
 
\affiliation{\aff{1}Physics of Fluids Group and Max Planck Center Twente, MESA+ Institute and J. M. Burgers Centre for Fluid Dynamics, University of Twente, P.O. Box 217, 7500AE Enschede, Netherlands
\aff{2}Faculty of Aerospace Engineering, Delft University of Technology, 2629 HS Delft, Netherlands
}

\begin{document}

\maketitle

\begin{abstract}
Turbulent convection systems are known to give rise to prominent large scale circulation. At the same time, the `background' (or `small-scale') turbulence is also highly relevant and e.g. carries the majority of the heat transport in the bulk of the flow. Here, we investigate how the small-scale turbulence is interlinked with the large-scale flow organization of Rayleigh-B\'enard convection. Our results are based on a numerical simulation at Rayleigh number  $Ra = 10^8$ in a large aspect ratio  ($\Gamma=32$) cell to ensure a distinct scale separation. We extract \emph{local} magnitudes and wavenumbers of small scale turbulence and find significant correlation of large scale variations in these quantities with the large-scale signal. Most notably, we find stronger temperature fluctuations and increased small scale transport (on the order of 10\% of the global Nusselt number $Nu$) in plume impacting regions and opposite trends in the plume emitting counterparts. This concerns wall distances up to $2\delta_\theta$ (thermal boundary layer thickness). Local wavenumbers are generally found to be higher on the plume emitting side compared to the impacting one. A second independent approach by means of conditional averages confirmed these findings and yields additional insight into the large-scale variation of small-scale properties. Our results have implications for modelling small-scale turbulence.
\end{abstract}

\section{Introduction}\label{sec:intro}
Natural convection is an important flow configuration with highly relevant applications in 
thermal convection in the atmosphere \citep{har01} and in the stars \citep{cat03} among many others. Fundamental aspects of natural convection are traditionally studied in the Rayleigh-B\'{e}nard  set-up, in which the flow evolves between parallel horizontal plates that are heated from below and cooled from above \citep{sig94, ahl09, loh10, chi12}. 
In this case, the strength of the thermal driving is expressed by the dimensionless Rayleigh number $Ra$, while, the dimensionless heat flux, which is the most important system response, is quantified by the Nusselt number $Nu$. 

A characteristic and very persistent feature of Rayleigh-B\'{e}nard convection (RBC) is the formation of a large-scale circulation (LSC). The LSC typically consists of one or several roll structures that fill the entire gap between the plates. Such structures continue to exist even in the highly turbulent state, at large $Ra$.
It has further been amply demonstrated \citep[e.g.][]{Fitzjarrald1976,har03,par04,har08,pan18,ste18,kru20} that the lateral extent of the LSC can reach multiple times the cell height $H$ in large aspect ratio domains. In laterally unconfined geometries, the LSC has therefore also been referred to as `superstructure' in the more recent literature. 

The existence of such large scale organization in an otherwise chaotic flow is arguably an exciting aspect of RBC, and is reminiscent of large-scale structures in other canonical flows \citep{hut07,Lee2018}. Their slow spatio-temporal evolution renders these structures prime candidates for flow modelling approaches. Such efforts are indeed already under way, e.g. by \citet{fon19} who used deep learning techniques to analyse superstructures up to moderate $Ra$ of $10^7$. However, these authors also acknowledged that the relative contribution of the LSC to the total heat transport decreases with increasing $Ra$. Similarly, \citet{kru20} found that at $Ra = 10^8$, superstructures account for a maximum of 30\% of the total heat transport. It is therefore clear, that the `background' turbulence (i.e. small to intermediate scales) plays an essential role that needs to be accounted for in low order models. This immediately raises the question on how structures of different sizes `interact' in RBC or more precisely: How does a certain large-scale state affect local properties of the small-scale field? This is exactly what is going to be addressed in this study. \rev{One should note, however, that the interactions investigated here are not the same as the interscale-transport of energy studied in e.g. \citet{Gayen2013}, \citet{tog15} or \citet{gre19}.}

Related findings on spatial variations in the small-scale statistics have recently been reported by \cite{He2019}. These authors found that in the center regions of the LSC (i.e.in the wind-shear regions) the decay of  temperature fluctuations away from the wall follows a power law dependence. Whereas in the plume emitting parts of the BL, temperature fluctuations drop of logarithmically with wall distance $z$. Relevant in the broader context are the studies by \cite{shi15} and \cite{Wang2016, Wang2018boundary}, who derived a BL equation for the mean and variance of the temperature profile, where they included the effect of the fluctuations into the otherwise laminar 2D BL equation. Further, \cite{Wang2019} disentangled temperature fluctuations throughout the RBC cell into homogeneous background (Gaussian) and non-homogeneous plume contributions (exponential).

\begin{figure}
\begin{center}
{\includegraphics[scale = 0.85]{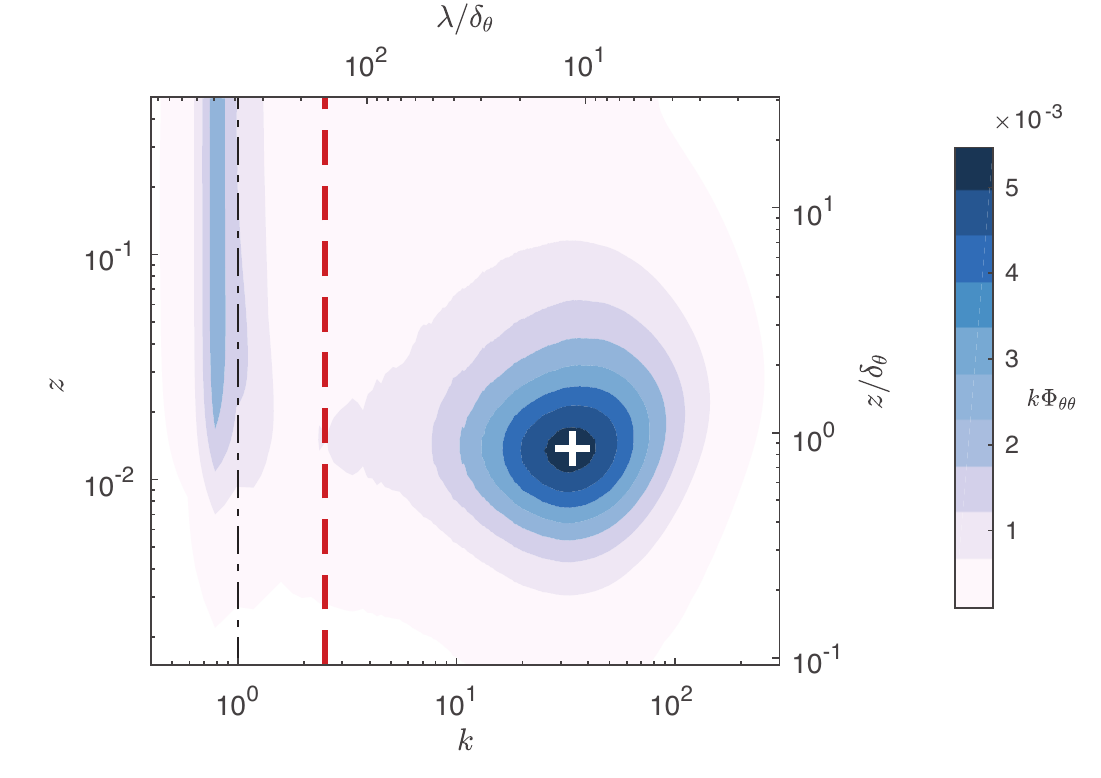}}
\caption{Premultiplied temperature spectrogram $k\Phi_{\theta \theta}$ at $Ra = 10^8$ and $\Gamma = 32$, where $k$ is the wavenumber and  $\lambda=(2\pi)/k$ the wavelength. The white cross located at $z = 0.85 \delta_\theta$ and $k = 34$ ($\lambda/\delta_\theta \approx 11.4$) indicates the location of the small-scale or `inner' peak. The black dashed-dotted line marks the superstructure scale $\hat{l}$ determined in \citet{kru20} based on the coherence metric and the red dashed line is located at the cutoff wavenumber $k_{\textrm{cut}} = 2.5$ used here.}
\label{fig:spec}
\end{center}
\end{figure}

The approach we choose here is to identify and to quantify modulation effects between large and small scales in terms of amplitude and wavenumber. Naturally, this first and foremost necessitates a clear definition of what those `large' (i.e. related to the LSC) and `small' (the rest) scales are. Such a distinction can readily be made for a superstructure configuration (aspect ratio $\Gamma = 32$) at $Ra  = 10^8$, as figure \ref{fig:spec} demonstrates. The premultiplied temperature spectrogram $k\Phi_{\theta \theta}$ shown there clearly separates into a large-scale peak at radial wavenumber $k \approx 1$ \rev{(normalized by $H$ as all length scales here)} and one at significantly smaller scales ($k \sim 10^2$). There is only little energy in the spectral gap between the two peaks, such that a scale decomposition can be achieved via spectral filtering. Here we chose a cut-off wavenumber $k_\mathrm{cut} = 2.5$ (indicated by the red line in the figure). \rev{Note, however, that the results are rather insensitive to the exact choice of $k_\mathrm{cut}$ as long as the value lies within the spectral gap, since in this case small and large scale signals are respectively dominated by energetic scales at wavenumber ranges far from $k_\mathrm{cut}$.} The normalized wavelength $\hat{l} = 2 \pi/\hat{k}$ \citep[where $\hat{k}$ is a characteristic  wavenumber of the LSC based on a coherence metric, see][]{kru20} of the large scale structures increases with increasing $Ra$ \citep{pan18,ste18,kru20}, while the structure size corresponding to the small scale peak scales with the thermal BL height $\delta_\theta$ \citep{kru20}. As demonstrated in \citet{Blass2020}, this leads to an increasing scale separation between the peaks  at higher $Ra$, rendering their interaction even more clear and relevant. 
\rev{It should be noted in this context that others have used spatial filtering \citep{gre19} or time-averaging \citep[e.g.][]{pan18,fon19} to extract the large-scale features in similar datasets.}

For the reasons outlined above, large aspect ratio cells are convenient for the present purpose and are therefore considered in the following. However, we note that there is little reason to believe that the effects studied are restricted to such domains. In fact, we expect scale interactions similar to those uncovered here to also play a role in smaller cells. Yet, spectral overlap of the peaks and other complications, e.g. due to corner vortices, render their analysis much more cumbersome in these cases.

Finally, we point out that a considerable number of studies on scale interactions in other flows exist, tracing back to \citet{brow77} and \citet{ban84}. Especially in the field of turbulent BLs, where a scale separation reminiscent of the situation displayed in figure \ref{fig:spec} occurs \citep{Hutchins2007large}, this topic has attracted considerable attention over the last decade \citep{mat09,mar10,chung:2010a,schlatter:2010ba,ber11,gan12,baars:2017a,dogan:2019a}. We will make use of the mature framework and tools developed in this context to quantify \textit{amplitude modulation} and \textit{wavenumber modulation} effects in RBC.

The paper is structured as follows; in \S\ref{sec:rbc} we present details of the numerical data set along with relevant parameters. We then quantify modulation effects using two independent analysis tools. First, using a wavelet-based method (\S\ref{ssec:red}--\S\ref{sec:correlations}) and then in \S\ref{sec:cond} on the basis of conditioned statistics. We conclude by discussing some physical implications of the results in \S\ref{sec:discussion}.

\section{Governing equations and parameters}
\label{sec:rbc}
\begin{table}
\centering
\begin{tabular}{c c c c| c c c c c c}
$Ra$ & \rev{$Pr$}& $N_x \times N_y \times N_z$ & $\Gamma$ & $t/T$ & $Nu$ & $\hat{l}$ & $\delta_\theta$ & $N_\theta$  \\
\hline
$1.0\times10^8$ & \rev{1}& $8192 \times 8192 \times 256$ & $32$ & $200$ & $30.94$ & $6.3$ & $0.016$  & $13$\\
\end{tabular}
\caption{Parameters of the dataset. Input parameters are the $Ra$ number, the numerical resolution in the horizontal ($N_x\times N_y$) and wall normal ($N_z$) directions, the aspect ratio $\Gamma$, an  the runtime $t$ of the DNS normalized by the free-fall timescale $T$. Relevant results are the $Nu$ number, the superstructure size $\hat{l}$ which is calculated from the coherence spectrum in \cite{kru20}, the thermal BL thickness $\delta_\theta$, and finally, the number of grid points that are in the thermal BL $N_\theta$.  } 
\label{tab:param}
\end{table}
The simulations used here have previously been reported in \citet{ste18} and we include some relevant details here for completeness.
The incompressible Navier Stokes (NS) equations within the Boussinesq approximation are solved using a second-order finite difference scheme. The dimensionless equations are:
\begin{equation}
\label{eq:NSkin}
\frac{\partial \tilde{\textbf{u}}}{\partial t} + \tilde{\textbf{u}}\cdot \nabla \tilde{\textbf{u}} = -\nabla \tilde{p} + \sqrt{\frac{Pr}{Ra}}\nabla^2 \tilde{\textbf{u}} + \tilde{\theta} \mathbf{e}_z,
\end{equation}
\begin{equation}
\label{eq:incom}
\nabla \cdot \tilde{\textbf{u}} = 0,
\end{equation}
\begin{equation}
\label{eq:NStherm}
\frac{\partial \tilde{\theta}}{\partial t} + \tilde{\textbf{u}} \cdot \nabla \tilde{\theta} = \frac{1}{\sqrt[]{Ra Pr}}\nabla^2 \tilde{\theta}.
\end{equation}
where the characteristic length scale is the height of the RBC cell $H$, the free-fall time scale $T= \sqrt{H/(g\alpha \Delta)}$ and the velocity scale is $\sqrt{g\alpha \Delta H}$, which is know as the free-fall velocity. Therein, $g$ is the gravitational acceleration (acting in opposite direction of the unit vector $\mathbf{e}_z$) , $\alpha$ is the thermal expansion coefficient and $\Delta$ is the temperature difference between the top and the bottom plate. Two dimensionless parameters appear after normalizing the  NS equations, namely the Prandtl number $Pr=\nu/\kappa$ and the Rayleigh number $Ra=\alpha g \Delta H^3/(\nu\kappa)$, where $\kappa$ is the thermal diffusivity and $\nu$ the momentum diffusivity (kinematic viscosity). Finally, the geometric control parameter is the aspect ratio $\Gamma=L/H$ of the square computational domain, with $L$ denoting the length of the cell in the periodic $x$ and $y$ directions. \rev{The simulations are performed with the second-order code AFiD \citep{vdPoel2015} and} parameters of the DNS are summarized in table \ref{tab:param}. Importantly, the BLs are  well resolved  and comply with the criterion derived in \cite{shi15}. The present dataset was originally used to confirm grid convergence in \citet{ste18} and is therefore even better resolved than the data reported there.

Instantaneous normalized values of pressure, temperature, and the velocity vector are denoted by $\tilde{p}$, $\tilde{\theta}$ and $\tilde{\mathbf{u}}$, respectively and we refer to the components of $\tilde{\mathbf{u}}$ as $\tilde{v}$ (horizontal) and $w$ (vertical). The operator $\langle \cdot \rangle$ is used to indicate averaging over the statistically homogeneous horizontal plane and time. We use the notation $\tilde{\xi}(x,y,z,t) = \Xi(z) +\xi(x,y,z,t)$ to decompose an arbitrary instantaneous quantity $\tilde{\xi}(x,y,z,t)$ into its mean $\Xi(z) = \langle \tilde{\xi} \rangle$ and a fluctuating part   $\xi(x,y,z,t)$.

\section{Results}\label{sec:results}

\begin{figure}
\begin{center}
{\includegraphics[width = \textwidth]{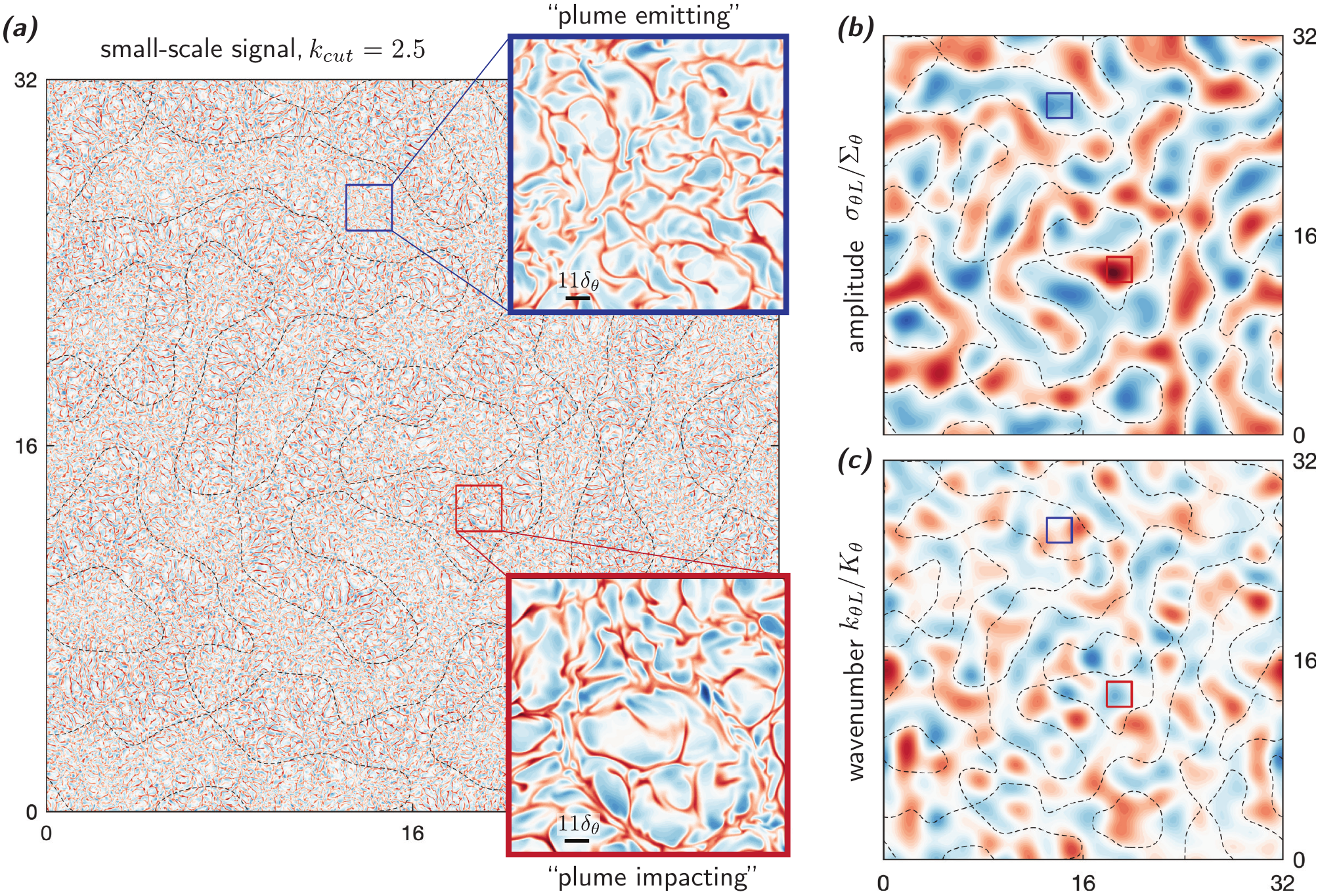}}
\caption{(a) Snapshot of the small-scale temperature fluctuations $\theta_S$ at a distance $z = \delta_\theta$ above the hot plate.  Blow-ups show plume emitting ($\theta_L>0$, blue box) and plume impacting ($\theta_L<0$, red box) regions. Panels (b) and (c) show the large-scale amplitude $\sigma_{\theta L}$ and wavenumber $\kappa_{\theta L}$, respectively for the snapshot shown in (a). The boxes indicate the positions of the blow ups in (a). Dashed black lines in all panels correspond to zero-crossings in the large-scale signal $\theta_L$ at mid-height $z = 0.5$. The scales of the color axes are $\pm0.36$, $\pm 0.2$ and $\pm 0.4 $ in (a), (b) and (c), respectively.}
\label{fig:fields}
\end{center}
\end{figure}

The essence of this study is best illustrated by the snapshot of the temperature field shown in figure \ref{fig:fields}a. There, the colour-contour represents $\theta_S(z=\delta_\theta)$, i.e. the small-scale temperature fluctuations at BL height. Note that we use subscripts $S,L$ to represent the small ($k>k_\mathrm{cut}$) and large-scale ($k<k_\mathrm{cut}$) signals, respectively\rev{, which are obtained from spectral filtering with cut-off wavenumber $k_\mathrm{cut}$}. Even though the large-scales ---and hence the effect of a direct \emph{superposition} of the LCS onto fluctuations at BL height observed before \citep{pan18,ste18,kru20}--- have been removed, there is still a distinct large-scale `organisation' visible for $\theta_S$. This pattern stems from  spatial variations of properties such as the \emph{local} amplitude and the \emph{local} wavenumber of $\theta_S$. Figure \ref{fig:fields}a further contains zero-crossings of $\theta_L$ (dashed lines), which make it obvious that these spatial variations in the small-scale signal are highly correlated to the large-scale dynamics.
In the next section we will introduce a procedure to capture the large-scale `footprint' in small scale turbulence by extracting local amplitude and wavenumber. On this basis we can then  quantify the correlation of this footprint with the large scales.

\subsection{Reduced-order representation of small-scale fluctuations}\label{ssec:red}
To quantify how the large-scale RB superstructures interplay with the small-scale turbulence ($\xi_S$), we have to resort to a simplified representation of the small-scale fluctuations ($\xi$ is a placeholder for the quantity concerned, \emph{e.g.} $\theta$, $w$, ... ). 
These fluctuations are broadband \citep[\emph{e.g.}][]{kru20} and pose a challenge to reduce the wealth of information in $\xi_S(x,y)$ to a comprehensible set of representative quantities. \rev{Here, this is achieved using a wavelet procedure, which allows to extract the large-scale spatial variations of the small-scale statistics. A similar procedure was applied to time series of velocity fluctuations by \cite{baa15}. However, their methodology is limited to one dimensional data and therefore needs to be extended to cope with two-dimensional (2D) fields to be applicable here.} Our procedure yields two new variables: (\emph{i}) the 2D space-varying magnitude $\sigma_\xi$ and (\emph{ii}) the 2D space-varying characteristic wavenumber $k_\xi$ of the energy in $\xi_S(x,y)$. Through correlating these two representative variables of the small-scale turbulence with the large-scale superstructures in $\xi_L$, we study the interaction between the two energetic regions that were identified in spectral space (see the discussion alongside figure~\ref{fig:spec}).

\begin{figure}
    \small
    \includegraphics[width = 1.00\textwidth]{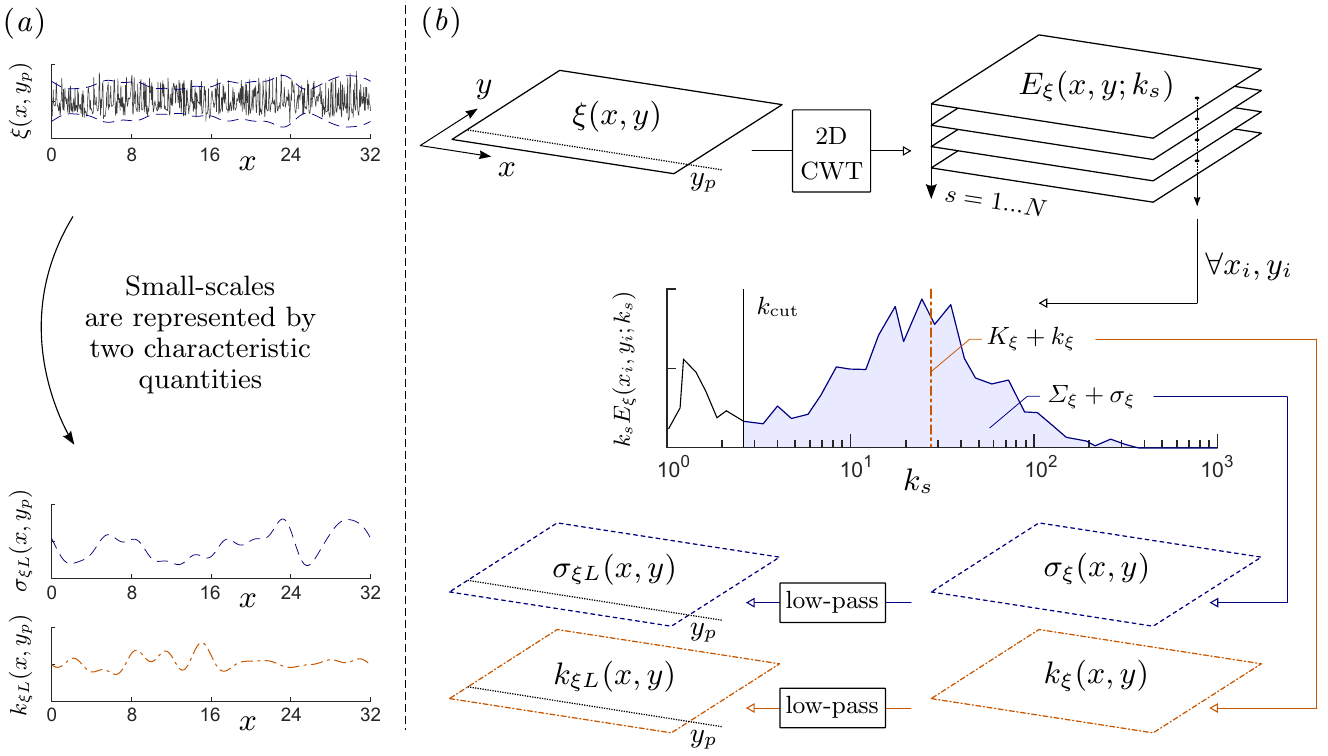}
    \caption{Diagram of the wavelet-based procedure, used in extracting local fluctuation properties of a turbulent quantity in a plane of RB turbulence, in particular the characteristic energy-amplitude of the fluctuations, $\sigma_{\xi L}(x,y)$, and a characteristic wavenumber of the fluctuations, $k_{\xi L}(x,y)$.}
    \label{fig:wavelet}
\end{figure}

Our procedure to obtain the reduced-order representation of the small-scale fluctuations is illustrated in figure~\ref{fig:wavelet}. A wall-parallel plane of $\xi$ forms the starting point in this diagram (top-left of figure~\ref{fig:wavelet}$b$). For ease of visualization, a one-dimensional subset taken along the $x$-direction of $\xi$ is shown in figure~\ref{fig:wavelet}($a$) for one particular $y = y_p$ location. Inspection yields a large-scale variation of the fluctuating magnitude $\sigma_{\xi L}$ (indicated by the dashed envelope) and a spatial variation in the characteristic wavenumber $k_{\xi L}$; this reflects the spatial variation in small scale properties already observed in figure \ref{fig:fields}a.  

\rev{The first step in providing a quantitative measure for these variations 
involves a two-dimensional (2D) wavelet transform. 
We restrict the discussion of this transform to a minimum here and refer to Appendix~A for details and to \emph{e.g.} \citet{far92} and \citet{add02} for further mathematical intricacies. The wavelet transformation is defined by a convolution of a mother wavelet $\Psi$ with $\xi(x,y)$, following 
\begin{equation}\label{eq:wavconv}
A_{\xi}(x,y;l_s) = \int\int \xi(x^*,y^*)\overline{\Psi}\left(\frac{x-x^*}{l_s},\frac{y-y^*}{l_s}\right){\rm d}x^*{\rm d}y^*,
\end{equation}
where the overline indicates the complex conjugate and $A$ is the so-called wavelet coefficient at a length scale $l_s$ of the self-similar wavelet. The wavelet $\Psi$ itself is a localized basis function, and is here taken as the isotropic Morlet wavelet (a 2D harmonic wave with a Gaussian envelope). This mother wavelet has no angular sensitivity, and is a suitable choice for RB turbulence, as there is no angular dependency in the statistics of this flow. The coefficient $A_{\xi}(x,y;l_s)$ is then a measure of the \emph{local} amplitude (at point $(x,y)$) at the scale $l_s$.}

\rev{A difficulty arises when relating wavelet decompositions to Fourier transforms because the wavelet scale $l_s$ does not have a precise counterpart in wavenumber space. This is because being localised in space, a wavelet $\Psi$ comprises energy at a range of wavenumbers instead of just a single one. The way to relate the wavelet scale to a wavenumber is then to determine the center wavenumber of the wavelet's Fourier transform, denoted as the pseudo-wavenumber $k_s$. We will adopt this concept here in order to better connect to other work on RBC, which predominantly employs Fourier transforms.
With this, coefficients $A_{\xi}(x,y;l_s)$ can be expressed as $A_{\xi}(x,y;k_s(l_s))$ and reflect the convolution magnitude as a function of a characteristic wavenumber. Finally, with the proper normalization, the square of the wavelet coefficients measures the energy level of a given field in terms of space and wavenumber, yielding the energy density
\begin{equation}
E_\xi(x,y;k_s) = \frac{A_{\xi}(x,y;k_s)^2}{k_s}.
\end{equation}}

The stack of $N$ planes in the top-right of figure~\ref{fig:wavelet}($b$) represents quantity $E_\xi(x,y;k_s)$ computed at $N$ different scales. For a given sample location, denoted as $(x_i,y_i)$, the local wavelet spectrum $E_\xi(x_i,y_i;k_s)$ is shown as a function of $k_s$. A total of $N = 33$ logarithmically spaced scales were used in the small-scale range $k_\mathrm{cut} < k < k_N$, with $k_N=4.5\times10^3$ denoting the DNS Nyquist wavenumber. The local wavelet spectrum is a measure of the energy spectrum at one spatial location. It is influenced by a domain surrounding $(x_i,y_i)$, {which is referred to as} the cone-of-influence of the wavelet, \emph{i.e.} the region spanned by the wavelet at a certain scale: a small-scale wavelet is more localized than a larger-scale wavelet. Since the boundaries of the computational domain are periodic, localized wavelet spectra exist for all points in the $(x,y)$ domain. When all localized spectra are averaged, one global wavelet spectrum is found for the full $(x,y)$ field which then matches the Fourier spectrum at one $z$ height in figure~\ref{fig:spec}. So, the key feature of $E_\xi(x,y;k_s)$ is that it preserves spatial information, which is absent in Fourier spectra.

So far, the decomposition of $\xi(x,y)$ into energy densities $E_\xi(x,y;k_s)$ at $N$ different scales has increased the dimensions of the system. 
For a reduced-order representation of the small-scale turbulence, we now define the following quantities: First we introduce a spatial field that captures the variation in the energy magnitude of $\xi(x,y)$. Integrating $E_\xi(x,y;k_s)$ over the small-scale wavenumber range $k_\mathrm{cut} < k < k_N$ and taking the square root yields  the space-varying standard deviation of $\xi_S(x,y)$ according to
\begin{equation}\label{eq:locals}
\Sigma_\xi + \sigma_\xi(x,y) = \left(\int_{k_\mathrm{cut}}^{k_N} E_\xi(x,y;k_s){\rm d} k_s\right)^{0.5}.
\end{equation}
Here, $\Sigma_\xi$ denotes the mean standard deviation of the small-scale {fluctuations (i.e. $\Sigma_\xi = \langle \xi_S^2 \rangle^{\frac{1}{2}}$)}, 
whereas $\sigma_\xi(x,y)$ is the zero-mean spatial fluctuation of the standard deviation. Only the large-scale variation of $\sigma_\xi$ is retained by low-wavenumber pass-filtering with $k < k_\mathrm{cut}$. The resulting spatial field $\sigma_{\xi L}(x,y)$ is interpreted as an envelope to the small-scale fluctuations and is plotted in figure~\ref{fig:wavelet}($a$). A second representative property of $\xi(x,y)$ is its local wavenumber, which we take as the first spectral moment of the local wavelet spectrum,
\begin{equation}\label{eq:localk}
K_\xi + k_\xi(x,y) = \frac{\int_{k_\mathrm{cut}}^{k_N} kE_\xi(x,y;k_s){\rm d} k_s}{\int_{k_\mathrm{cut}}^{k_N} E_\xi(x,y;k_s){\rm d} k_s} = \frac{1}{\left(\Sigma_\xi + \sigma_\xi(x,y)\right)^2} \int_{k_\mathrm{cut}}^{k_N} kE_\xi(x,y;k_s){\rm d} k_s.
\end{equation}
Similarly as for the standard deviation, the mean wavenumber of the small-scale energy in $\xi(x,y)$ is $K_\xi$ and its zero-mean fluctuations are denoted as $k_\xi(x,y)$.  Low-wavenumber pass-filtering the latter yields $k_{\xi L}(x,y)$. In summary, the raw small-scale field $\xi_S(x,y)$ is represented by $\sigma_{\xi L}(x,y)$, which reflects the large-scale spatial deviation of the energy (or intensity) in the small-scale fluctuations from its mean energy. Further $k_{\xi L}(x,y)$ describes the deviation of characteristic local wavenumber from the mean wavenumber. Thus, positive (negative) $k_{\xi L}$ means that the small-scale fluctuations are of a higher (lower) wavenumber locally, compared to an averaged over the entire domain.  Results of $\sigma_{\theta L}$ and $k_{\theta L}$ computed for the snapshot in figure \ref{fig:fields}a are included in figure \ref{fig:fields}b,c, respectively. To what degree these quantities are correlated  with the large-scale signal can be gauged by comparing to the zero-crossings of $\theta_L$ (shown as black lines). We will quantify  this aspect next.

\subsection{Correlations of large and small scales}
\label{sec:correlations}
\begin{figure}
\begin{center}
{\includegraphics[width = \textwidth]{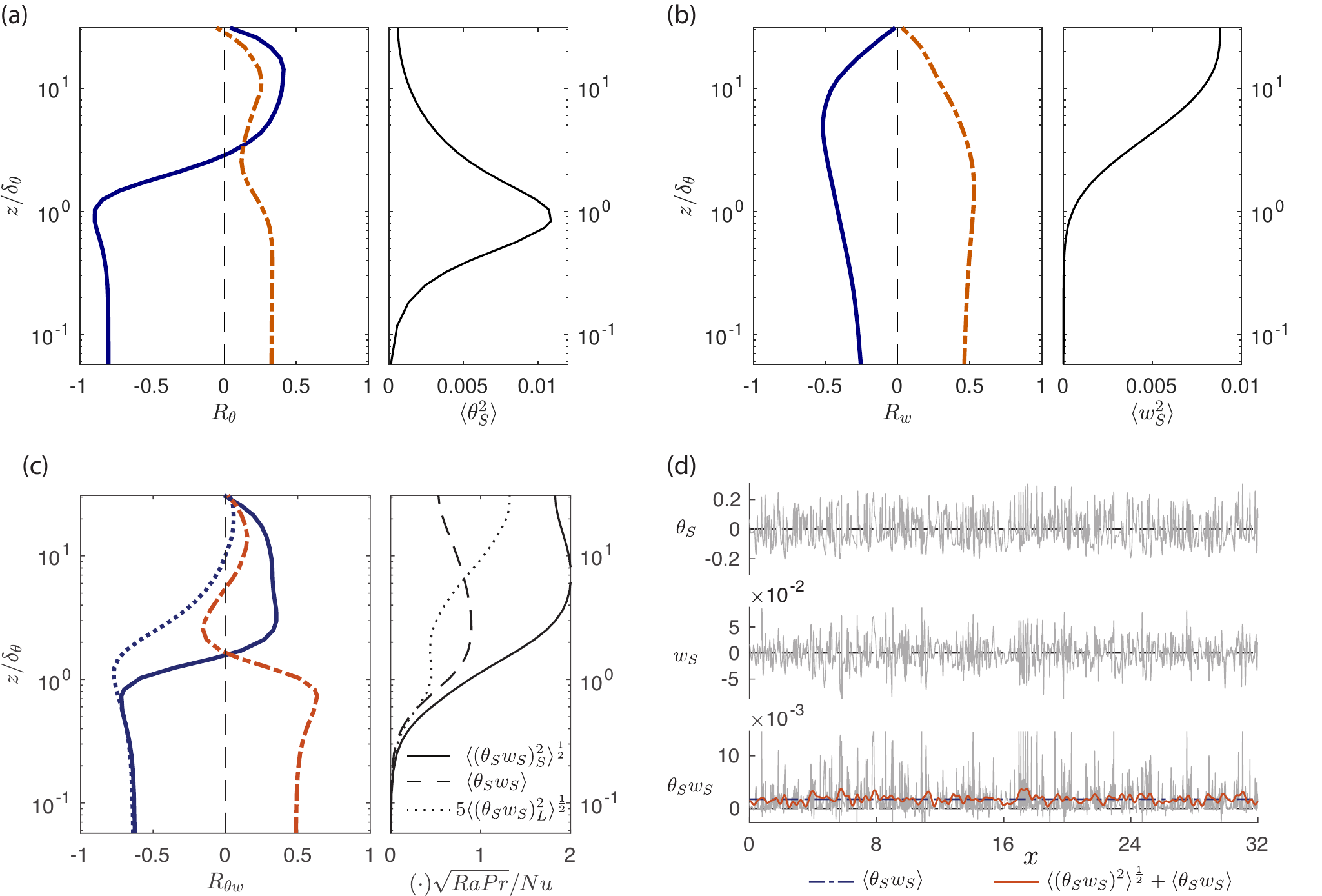}}
\caption{Wall-normal profiles of the amplitude modulation coefficient (solid dark blue), the wavenumber modulation coefficient (dash-dotted orange) and the small-scale energy $\overline{\sigma_\psi}^2$ (solid black) for (a) the temperature $\theta$, (b) the wall normal velocity $w$ and (c) the small-scale turbulent heat transport term $(\theta_S w_S)$. The blue dotted line in (c) is the correlation coefficient for $(\theta_S w_S)_L$. Panel (d) illustrates the construction of the $(\theta_S w_S)$ signal. It contains a large-scale part (solid red), a non zero-mean (dashed black line) and the remaining small-scale signal, of which the respective variations with $z$ are shown in the right plot of panel (c).}
\label{fig:AM}
\end{center}
\end{figure}
To study the influence of the large scales on the small scales in RBC, \rev{we define an amplitude modulation coefficient similar to what is used in work on turbulent boundary layers \citep[e.g.][]{ban84,mat09}.} This coefficient is the normalized correlation between the large-scale envelope of the small-scale turbulence $\sigma_{\xi,L}(x,y,z)$ and the large-scale signal of the temperature at mid-height $\theta_L(x,y,H/2)$,
\begin{equation}\label{eq:AMcoef}
R_{\xi,\sigma}(z) = \frac{\langle \sigma_{\xi L} (x,y,z) \theta_L(x,y,H/2) \rangle}{\sqrt{\langle \sigma_{\xi L}(x,y,z)^2\rangle}  \sqrt{\langle \theta_L(x,y,H/2)^2\rangle}}.
\end{equation}
It is important to emphasize that in this definition, we always use the large-scale temperature signal at mid-height $\theta_L(x,y,H/2)$ as reference. Clearly, other choices are possible here. However, since the LSC is known to be highly coherent along the vertical direction and because $\theta_L$ correlates almost perfectly with the large-scale velocity signal \citep{kru20}, this will have no significant impact on our results and conclusions. By definition, $R_{\xi,\sigma}$ varies between $-1$ (perfect anti-correlation) and $1$ (perfect correlation and $R_{\xi,\sigma} = 0$ implies uncorrelated signals.   

Analogously, the wavenumber modulation coefficient $R_{k,\xi}(z)$ is defined by correlating the large-scale part of the wavenumber fluctuations $k_{\xi L}(x,y,z)$ to $\theta_L(x,y,H/2)$,
\begin{equation}\label{eq:WMcoef}
R_{\xi,k}(z) = \frac{\langle k_{\xi L} (x,y,z) \theta_L(x,y,H/2) \rangle}{\sqrt{\langle k_{\xi L}(x,y,z)^2 \rangle}  \sqrt{\langle \theta_L(x,y,H/2)^2\rangle}}.
\end{equation}
We note that (\ref{eq:AMcoef}) and (\ref{eq:WMcoef}) define a \emph{normalized} correlation. The physical relevance of such a correlation, however, also depends on how much (absolute) energy is contained in the fields of $\sigma_{\xi,L}$ and $k_{\xi L}$. For the snapshot in figure \ref{fig:fields}, fluctuations in both $\sigma_{\theta,L}$ and $k_{\theta L}$ are on the same order (tens of percent) as their respective means and hence significant. We will revisit this point in \S\ref{sec:cond} in more detail.

Results for both modulation coefficients in the temperature field are presented as a function of the height $z$ in figure \ref{fig:AM}a. The most striking observation from this plot is that the small-scale amplitude of $\theta$ is highly correlated to the large scales up to a distance of about $2\delta_\theta$ above the plate. The peak of this correlation at which remarkably $R_{\theta,\sigma}<-0.9$ is found at $z\approx \delta_\theta$. This coincides with the location of the strongest small-scale fluctuations in the temperature field as is obvious from the right-hand panel of figure \ref{fig:AM}(a) as well as from the spectral peak of the energetic small-scales in figure \ref{fig:spec}. 
Throughout this paper, all statistics are plotted with reference to the hot plate. Hence, the negative sign of $R_{\sigma,\theta}$ in the near-wall region implies that in plume impacting regions (where $\theta_L<0$) the local small-scale energy is higher ($\sigma_{\theta L}>0$). This is also confirmed visually from the blow-ups in figure \ref{fig:fields}a.

Interestingly, $R_{\theta,\sigma}$ crosses zero around $z=2\delta_\theta$ and attains somewhat lower, yet distinctly non-zero values around $R_{\theta,\sigma}\approx 0.4$ in the bulk. The subsequent drop to zero at mid-height is a symmetry requirement. The crossover height of $z=2\delta_\theta$ is consistent with recent findings by \cite{He2019}, who found similar levels of temperature fluctuations in plume emitting and sheared regions at this height. 
Further away from the wall, \cite{He2019} found that plume emitting regions contain significantly higher fluctuations of $\theta$, which is also consistent with the present observations.

Compared to $R_{\theta,\sigma}$, the wavenumber modulation coefficient for $\theta$ (dashed red line) remains relatively low at all $z$. Also in this case, the strongest correlation encountered within the thermal BL with $R_{\theta,k}\approx 0.3$. Unlike $R_{\theta,\sigma}$, $R_{\theta,k}$ does not change sign and remains positive throughout, which implies that structures in the temperature field are slightly more `compact' in plume emitting regions.  

In principle, correlation coefficients can be evaluated for all fluctuating quantities. Here we will in addition to $\theta$ only consider the wall-normal velocity $w$. This is motivated by the fact that $w$ also features in the turbulent heat transport $\theta w$.  
From figure \ref{fig:AM}(b), we find that the amplitude modulation coefficient for $w$ is negative at all heights at intermediate correlation levels of about $0.4$. On the contrary, $R_{w,k}$ is positive at all $z$ locations and reaches values up to 0.5. Both, $R_{w,\sigma}$ and $R_{w,k}$ peak in the range 1-10 $\delta_\theta$, over which also the fluctuation level in $w_S$ builds up to its bulk value (see right-hand panel in figure \ref{fig:AM}b). The picture that emerges here is that $w_S$ fluctuations are of lower magnitude but higher wavenumber in plume emitting regions. 

Besides the individual fields of $w_S$ and $\theta_S$, it is meaningful to study the modulation effects for the turbulent heat transport $\theta w$. Initially, we apply the same analysis leading to (\ref{eq:AMcoef}) and (\ref{eq:WMcoef}) also to 
 $\theta w$. The corresponding results, $R_{\theta w,\sigma}$ (solid blue line) and $R_{\theta w,k}$ (dashed red), characterize zero-mean fluctuations of $(\theta w)_S$ and are shown in figure \ref{fig:AM}c. Both display substantial correlation with the large scales up to BL height, at which both the mean and the fluctuations in the normalised small-scale heat transport already reach the level of $Nu$, as can be seen from the right panel in \ref{fig:AM}c. Consistent with observations for $\theta$ and $w$ individually, we find $R_{\theta w,\sigma}<0$ and $R_{\theta w,k}>0$ for $z\lessapprox \delta_\theta$, while at larger wall distances $R_{\theta w,\sigma}$ is slightly positive with no significant wavenumber modulation.
 
There is, however, another point to be made here. This relates to the fact that even though individually $\theta_S$ and $w_S$ are zero-mean and small-scale only, their product $\theta_Sw_S$ contains not only a significant mean (that contributes the bulk of the total heat transport above $z \approx \delta_\theta$, see figure \ref{fig:AM}c, right panel), but also large scale fluctuations. This is illustrated for a 1-D sample in figure \ref{fig:AM}d. 
The large scale variations $(\theta_S w_S)_L$ are not equally strong as those at smaller scales (note the factor $5$ for the rms of $(\theta_S w_S)_L$ in figure \ref{fig:AM}c), yet their standard deviation still amounts to about 10\% of $Nu$ around BL height which is significant. At this wall distance, $(\theta_S w_S)_L$ displays a pronounced negative correlation with $\theta_L$ (dotted blue line in figure \ref{fig:AM}c), which extends even somewhat deeper into the bulk compared to $R_{\theta w,\sigma}$. This implies that the local heat transport carried by small-scale fluctuations is locally reduced (increased) in plume emitting (impacting) regions around BL height.

Correlation coefficients are very useful in quantifying the overall nature of scale interactions. However, they do not reveal all details about the spatial organisation in the flow. We will therefore complement the results discussed so far by additionally considering conditioned averages in the following.

\subsection{Conditioned averages}\label{sec:cond}
\subsubsection{Averaging procedure}
\begin{figure}
\begin{center}
{\includegraphics[width = \textwidth]{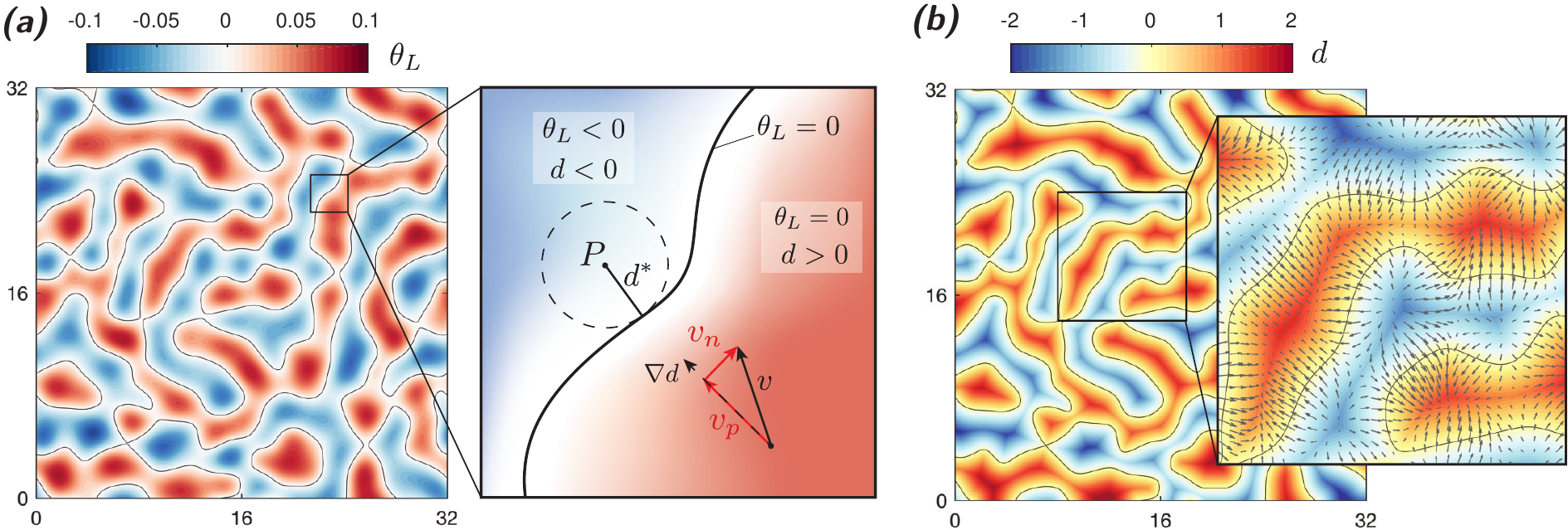}}
\caption{Illustration of the definition of the conditioned average. (a) Large-scale part of the temperature fluctuations at mid-height $\theta_L(z = 0.5)$. Black lines in all panels are iso-contours of $\theta_L =0$. Every point $P$ in the plane is assigned a distance $d$, where the magnitude of $d$ is determined by the distance to the nearest point on the iso-contour (see blow-up in (a)) and the sign is determined by that of $\theta_L$ at  point $P$ . The resulting $d$-field corresponding to the snapshot of $\theta_L$ in (a) is shown in panel (b). The in-plane velocity vector at each point (see blow-up in (b) is decomposed into a component along ($v_p$) and normal ($v_n$) to the gradient $\nabla d$ as shown in the blow-up in (a).}
\label{fig:balldist}
\end{center}
\end{figure}
The results obtained so far indicate that the properties of the small-scale turbulence, especially in the vicinity of the BLs, vary significantly depending on the location within the LSC. In this section, we aim to investigate those spatial variations in more detail. To that end, we employ conditioned averages with respect to the zero crossing of the large-scale signal $\theta_L$. The definition of the new conditioned coordinate $d$ is illustrated in figure \ref{fig:balldist}(a). For every point $P$ in a plane $d^*$ measures the smallest distance to a contour of $\theta_L =0$, i.e. $d^*$ is the smallest radius of a circle around $P$ that just touches a point on the zero-contour. To distinguish large-scale positive from large-scale negative regions, $d$ then takes the sign of $\theta_L$, such that
\begin{equation}
d=\textrm{sign}(\theta_L)d^*.
\end{equation}
 The  $d$-field corresponding to the $\theta_L$ snapshot in figure \ref{fig:balldist}(a) is shown in panel (b) of the same figure. The blow-up in figure \ref{fig:balldist}b presents the large-scale horizontal velocity vector at BL height (with the same cut-off wavenumber as $\theta_L$). This plot demonstrates that the LSC flow at the hot plate is generally directed from small $d$ to large $d$. The largest large-scale velocity magnitudes are encountered around $d\approx 0$. Since the large scale circulation is approximately aligned with $\nabla d$, it is useful to decompose the horizontal velocity vector $\textbf{v}$ into a component $v_p$ along $\nabla d$ and one ($v_n$) normal to it as illustrated in the blow-up of figure \ref{fig:balldist}a. 

To retain the large-scale organization in the flow, we then compute conditional averages by taking the mean overall points with constant $d$. Mathematically, this results in a triple decomposition for any instantaneous quantity $\tilde{\xi}(x,y,z,t)$ according to
\begin{equation}
\tilde{\xi}(x,y,z,t) = \Xi(z)+ \overline{\xi}(z,d) + \xi'(x,y,z,t),
\end{equation} 
where the overline denotes the conditional average, and $\xi'(x,y,z,t)$ is the fluctuation about the conditional and the temporal average. 

\begin{figure}
\begin{center}
{\includegraphics[width = \textwidth]{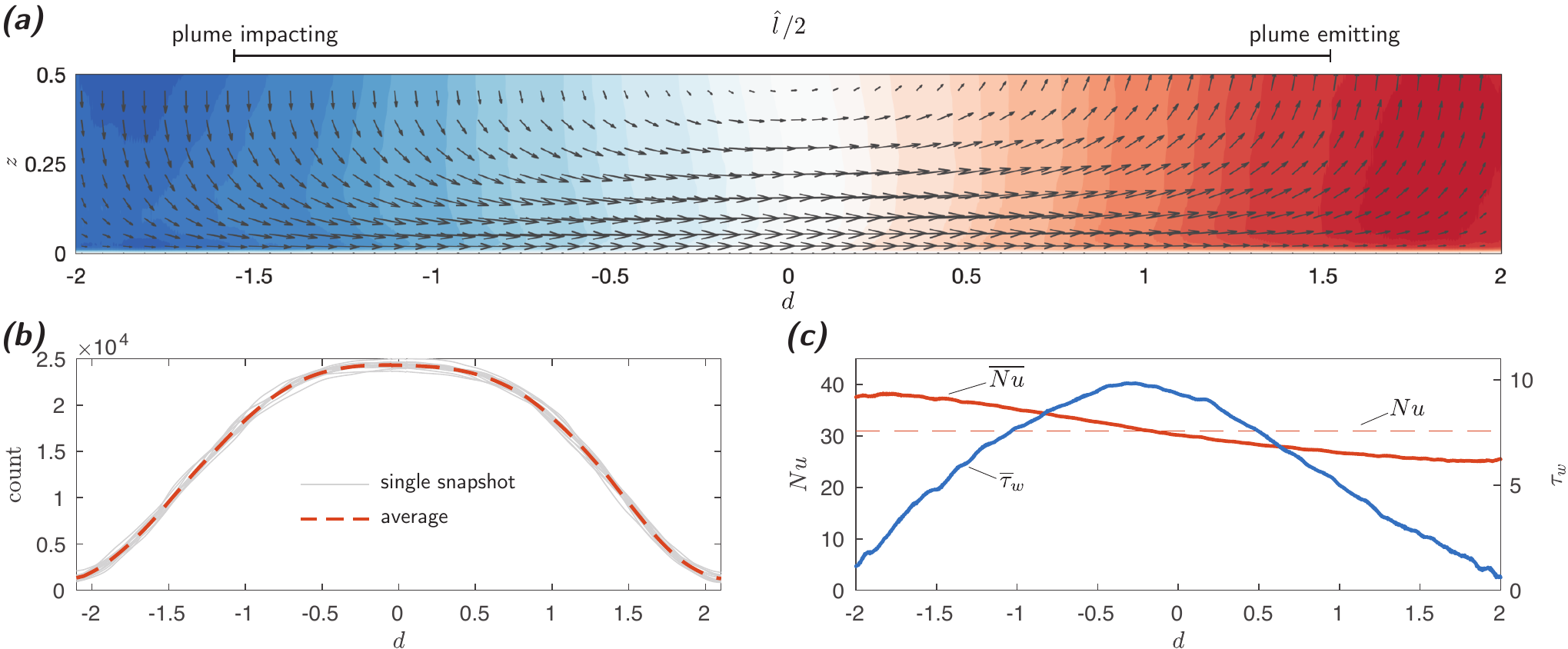}}
\caption{(a) Superstructure convection roll as obtained from the conditional averaging procedure. The color contours represent $\overline{\theta}$ and span the range $\pm 0.1$. Vectors indicate the field $(\overline{v}_p, \overline{w})$ with 15 data points skipped between adjacent vectors. (b) Number of points in a single plane as a function of the conditioning parameter $d$ for a total of $13$ snapshots.  (c) Variation of the conditioned Nusselt number $\overline{Nu}$ and of the conditioned shear stress $\overline{\tau}_w$  as functions of $d$.}
\label{fig:condavg}
\end{center}
\end{figure}

Conceptually, the conditioned means $\overline{\xi}(z,d)$ correspond to the large-scale fluctuation field and represent  the LSC, while the fluctuations $\xi'(x,y,z,t)$ around the conditioned mean are related to small-scale turbulence. It is hence  of interest to analyse the spatial distribution of statistics related to $\xi'$ in the present context. 
Before doing so, we briefly  present some results for the conditioned mean in order to orient the reader about basic  properties of the LSC in the present configuration. For that purpose, the contour of the conditionally averaged temperature $\overline{\theta}$ is shown in figure \ref{fig:condavg}(a) overlaid with  vectors representing $(\overline{v}_p, \overline{w})$. It is clear from this figure that the ranges $d<-1$ and $d>1$ correspond to the plume impacting and emitting zones of the LSC, respectively, and the strongest horizontal flow is encountered for $-1<d<1$. Further, the histogram in figure \ref{fig:condavg}(b) shows that the distribution of $d$ is already well-converged from a single snapshot only with only few counts beyond half the typical superstructure size $\hat{l} \approx 3$ for this case. Relevant transport quantities that can be derived from the fields shown in figure \ref{fig:condavg}(a) are the local wall shear stress $\overline{\tau}_w = |(\partial_z \overline{v_p}|_{z=0})|$, which is seen to peak at slightly negative $d$ (see figure \ref{fig:condavg}c). Moreover, we observe a significant spatial variation of $\pm 0.3Nu$ for the local heat transport $\overline{Nu} =  \partial_z(\Theta +\overline{\theta})|_{z=0}$. As  figure \ref{fig:condavg}c shows, heat transport is  enhanced in plume impacting regions and decreased in plume emitting regions. These observations are similar to those reported for RBC at $\Gamma=1$, $Pr=0.79$ and matching $Ra$, see \cite{wag12}. For a more detailed study on large-scale properties including their $Ra$ dependence, we refer to \citet{Blass2020}.

\subsubsection{Conditioned variance}
\begin{figure}
\begin{center}
{\includegraphics[width = \textwidth]{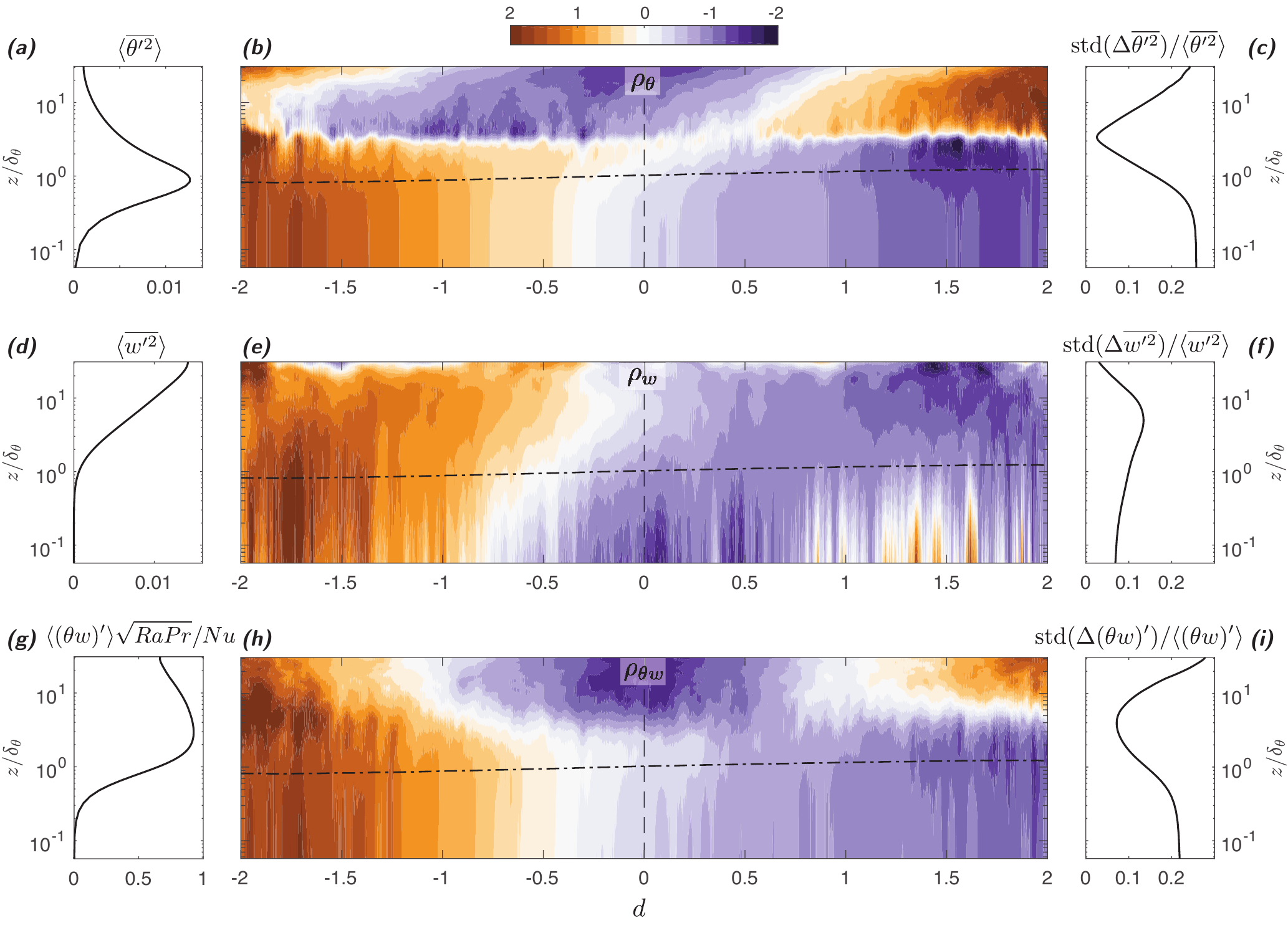}}
\caption{Statistics of the conditioned fluctuations of $\theta$ (a-c) and $w$ (d-f). Additionally, statistics of $(\theta w)'$ are shown in panels (g-i). Note that the standard deviation operator used in (c,f,i) implies horizontal averaging only, i.e. $\textrm{std}(\cdot) =  \langle (\cdot)^2 \rangle^{1/2}$ . The horizontal dashed line in (b,e,h) shows the local thermal BL height $\overline{\delta}_\theta(d) = 1/(2 \overline{Nu})$. The colorbar applies to all contour plots. }
\label{fig:condavg2}
\end{center}
\end{figure}

Several statistical properties are relevant when examining the fields of conditioned fluctuations $\xi'$ in terms of their magnitude. The most basic one is the variance $\langle \xi'^2\rangle (z)$, which indicates how energetic the small scales are at a given height (note that $\langle \cdot \rangle$ continues to imply horizontal averaging, in this case over $d$). The small-scale temperature variance $\langle \overline{\theta'^2}\rangle (z)$ is shown in \ref{fig:condavg2}a and indeed resembles $\langle \theta_S^2\rangle$ in figure \ref{fig:AM}a closely. The main interest here is in lateral (along $d$) variations of the variance, which we highlight by considering the difference 
$\Delta \overline{\xi'^{2} }(z,d) = \overline{\xi'^{2}}(z,d) - \langle \overline{\xi'^{ 2}}\rangle (z)$. This quantity is akin to the large scale envelope $\sigma_{\theta L}$.
 We plot the normalized standard deviation of $\Delta \overline{\theta'^{2} }$ in figure \ref{fig:condavg2}c as a measure for how much the temperature variance varies with $d$ at a given height.
Finally, we define the coefficient
\begin{equation}
\rho_\xi(z,d) \equiv \frac{\overline{\xi'^{2}}(z,d) - \langle \overline{\xi'^{ 2}}\rangle(z)}
{\langle (\overline{\xi'^{2}}(z,d) - \langle \overline{\xi'^{2}}\rangle(z))^2\rangle^{1/2}} =
\frac{\Delta \overline{\xi'^{2}}(z,d)  }{\mathrm{std}\left( \Delta \overline{\xi'^{2} }\right)(z)},
\label{eq:cond_var}
\end{equation}
to illustrate how $\Delta \overline{\xi'^{2} }$ varies across the LSC. Results for $\rho_\theta$ are displayed in figure \ref{fig:condavg2}b. Entirely consistent with figure \ref{fig:AM}(a), also this plot indicates enhanced (reduced) small-scale temperature fluctuations in the impacting (emitting) region below $z \approx 2\delta_\theta$. At this height, both  $\langle \overline{\theta'^2} \rangle$  and $\Delta \overline{\theta'^{2} }$ are of significant magnitude. Also the change in sign for $R_{\theta \sigma}$ towards positive values observed for $z \gtrsim 2\delta_\theta$ is echoed in the distribution of $\rho_\theta$. Generally, $\rho_\theta >0$ at $d>0$ (and vice versa) in the bulk. However, the zero-crossing clearly does not coincide with $d =0$ in this case, which is testament to the additional information provided by considering the conditioned averages.

Results analogous to those presented  for the temperature field  in figure \ref{fig:AM}a-c are included for $w$ in panels d-f of the same figure. Also in this case, there is a striking agreement between the distribution of $\rho_w$, which indicates stronger fluctuations on the plume impacting side, and  $R_{w \sigma}$ in figure \ref{fig:AM}b. 

Finally, we turn again to the turbulent heat transport. Here, we define the small-scale contribution according to
\begin{equation}
(\theta w)'(z,d) = \overline{\theta w}(z,d) - \overline{\theta}\overline{w}(z,d),
\label{eq:def_trans}
\end{equation}
that is, we subtract the part of the heat transport carried by the LSC from the overall turbulent transport. As figure \ref{fig:condavg2}g shows, $\langle (\theta w)' \rangle $ carries a bulk of the global transport at wall distances beyond the thermal BL height (note also the agreement with $\langle \theta_S w_S \rangle$ in figure \ref{fig:AM}c). The corresponding coefficient 
\begin{equation}
\rho_{\theta w}(z,d)  \equiv \frac{(\theta w)'(z,d) - \langle  (\theta w)' \rangle (z)}
{\langle ((\theta w)'(z,d) - \langle  (\theta w)' \rangle_d(z))^2\rangle^{1/2}} =
\frac{\Delta (\theta w)'(z,d)}{\mathrm{std}\left( (\theta w)' \right)(z)},
\label{eq:cond_trans}
\end{equation}
is analogous to (\ref{eq:cond_var}) but provided explicitly for clarity. Importantly, it is clear from (\ref{eq:def_trans}) and  (\ref{eq:cond_trans}) that $(\theta w)'$ and hence also  $\rho_{\theta w}$ describe how the net heat transport by the small scales varies spatially. The results presented here therefore approximately correspond  to $(\theta_S w_S)_L$ in figure \ref{fig:AM}c. The contour plot of $\rho_{\theta w}$ (figure \ref{fig:condavg2}h)  then also confirms the observation that up to several BL heights, small-scale heat transport is stronger in plume impacting regions. In the bulk, the correlation coefficient for $(\theta_S w_S)_L$ is essentially zero (see dotted blue line in figure \ref{fig:AM}c). However, this does not imply that the small scale heat transport is uniform there: As figure \ref{fig:condavg2}h reveals, small-scale transport is enhanced for both small and large $d$ and lower in the center of the LSC around $d=0$. Even though a pronounced large scale organisation evidently exists, this pattern does not lead to a non-zero correlation coefficient due to its reflection symmetry around $d=0$. The  horizontal standard deviation  $\Delta (\theta w)'$ amounts to typically 10\% to 20\% of $\langle (\theta w )\rangle$ (see figure \ref{fig:condavg2}i). These figures are comparable in terms of $Nu$ (cf. figure \ref{fig:condavg2}g), such that spatial variations of the small scale heat transport are indeed significant on a global level.

\subsubsection{Conditioned wavenumber}
\begin{figure}
\begin{center}
{\includegraphics[width = \textwidth,trim={0 0cm  0 0},clip]{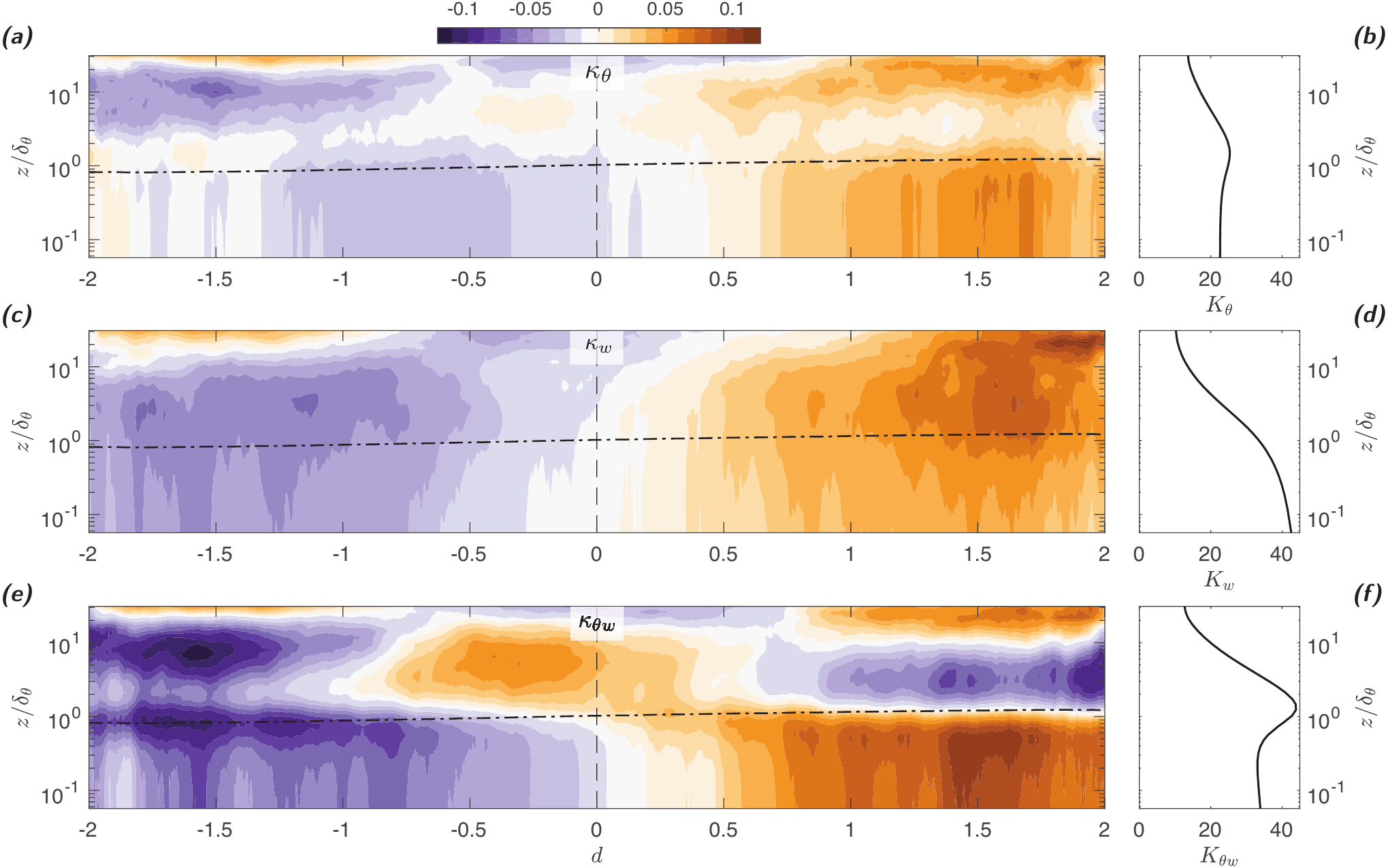}}
\caption{Normalized conditioned wavenumber fluctuations of different fields: (a) temperature, (c) wall normal velocity and (e) heat transport. The respective means as a function of $z$ used for normalization are shown in (b,d,f). The colorbar applies to all contour plots.}
\label{fig:condavg_freq}
\end{center}
\end{figure}

As a final point in our analysis, we reconsider spatial variations of the local wavenumber defined in (\ref{eq:localk}). We normalize the conditioned wavenumber fluctuations $\overline{k}_\xi(z,d)$ by their mean $K_\xi (z)$ at the respective height to obtain
\begin{equation}
\kappa_\xi(z,d) = \frac{\overline{k}_\xi(z,d)}{K_\xi (z)}.
\end{equation}
The results for $\kappa_\theta$, $\kappa_w$, and $\kappa_{\theta w}$ are presented in figure \ref{fig:condavg_freq}(a,c,e), respectively,  along with the corresponding means $K_\theta$, $K_w$, and $K_{\theta w}$ in figure \ref{fig:condavg_freq}(b,d,f) for reference. 

Largely, the plots of  $\kappa_\theta$, $\kappa_w$ recount the observations made in figure \ref{fig:AM}(a,b), indicating higher wavenumbers on the plume emitting side for both quantities. This trend appears to be slightly more pronounced for $w$ compared to $\theta$. 

The strongest wavenumber modulation is observed for  $\kappa_{\theta w}$ in figure \ref{fig:condavg_freq}e. The trend there agrees with $\kappa_\theta$ and  $\kappa_w$ (and also with $R_{\theta w,k}$ in figure \ref{fig:AM}c) in indicating higher wavenumbers at $d>0$ approximately  up to BL height $\delta_\theta$. Similar to $\rho_{\theta w}$, in the bulk a pronounced modulation pattern is also visible for $\kappa_{\theta w}$, which remains hidden in $R_{\theta w,k}$ due to its symmetry. In the case of $\kappa_{\theta w}$, this pattern is reversed for $\delta_\theta \lessapprox z \lessapprox 10\delta_\theta$, for which positive $\kappa_{\theta w}$ is found in the center instead of at large $|d|$.

\section{Concluding remarks}
Our analysis on the basis of  correlations as well as through conditioned averaging has revealed significant large-scale organization in small-scale turbulence fields in RBC. These modulation effects were seen to be strongest within ---but by no means limited to--- a layer close to the wall. Changes in the small-scale organization patterns occurred around a wall height of $z \approx 2\delta_\theta$. \rev{At this wall distance, the mean temperature profile approaches the bulk value and the location also coincides with the boundary of the `mixing zone' described by \cite{Wang2019}.}
 They find that this mixing zone is characterized by the emergence of intermittent, strong, hot plumes, that lead to long exponential tails in the probability distribution function (PDF) of the temperature. In their case, the mixing region differs from the `boundary region', where the temperature PDF is described by homogeneous, purely Gaussian, background fluctuations. This distinction does not appear to be relevant in the present context. 
 Further, the crossover height of $z \approx 2\delta_\theta$ links to a sharp decrease in vertical coherence with reference at BL height of the temperature field at large wavenumbers observed by \cite{kru20}. This is indicative that indeed small structures emanating from the BL do not extend beyond this height, which may explain why the small-scale organisation changes there. 
 
\begin{figure}
\small
\begin{center}
{\includegraphics[draft = false,width = \textwidth]{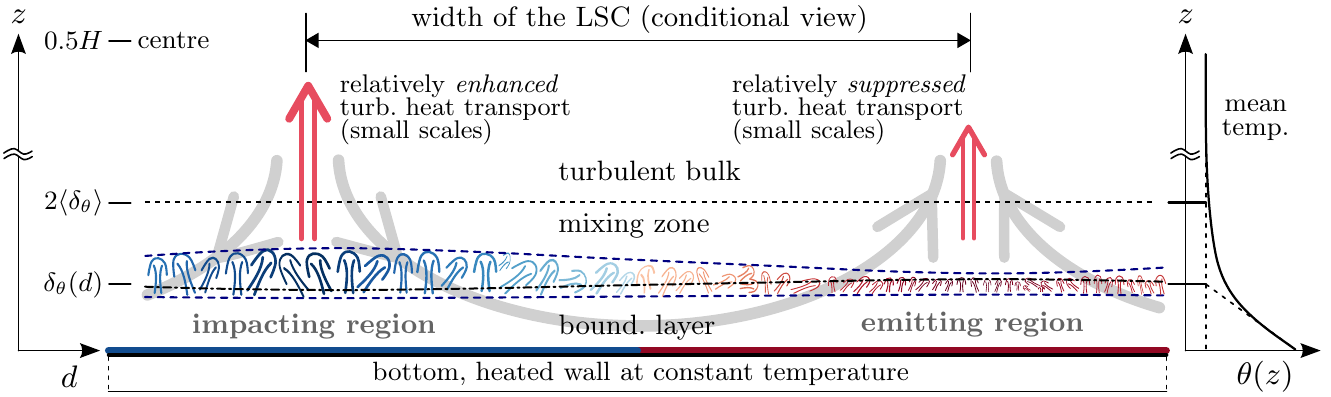}}
\caption{Schematic of the large-scale footprint in small-scale near wall RBC turbulence. Small-scale structures are shown as plumes with their height indicating the local magnitude and variations in spacing the local size (i.e. the wavenumber) of the fluctuations. \rev{The qualitative mean temperature profile serves to put the different regions into perspective.}}
\label{fig:illustration}
\end{center}
\end{figure}

 In RBC, the small-scales are typically most energetic close to the wall, making our results most relevant in this region; the strongest correlations are also encountered here with coefficients up to 0.9 for the temperature amplitude. Our findings regarding the near-wall zone are summarized graphically in figure \ref{fig:illustration}. Up to a wall distance of $z \approx 2\delta_\theta$, we find that the amplitude of small scale structures (indicted as plumes in figure \ref{fig:illustration}) in the plume impacting region is higher, while the wavenumber (and hence the structure size) is higher there. Most importantly, we find that the turbulent heat transport carried by the small-scales is higher (on the order of 10\% of $Nu$) in the plume impacting region and decreased by the same amount on the emitting side.
Significant large-scale organisation of the small scales was also identified in the turbulent bulk of the flow. However, the patterns are more complicated in this case and vary between different quantities, which precludes a straightforward (graphical) summary of the findings. Among the most notable results concerning the bulk flow is the fact that small-scale transport is enhanced at both sides of the LSC and is minimal in the center around $d = 0$.

\label{sec:discussion}

Thus far, we have not touched on the cause of the spatial variations in the small-scale statistics. The vertical localisation of the peak in $\langle \theta_S^2\rangle$ correlates with the local maximum in the production term  of temperature variance, which is proportional to $\theta w \,\partial_z \Theta$ \citep[see e.g.][]{Deardorff1967,tog15}. It appears therefore likely that also horizontal differences in $\langle \theta_S^2\rangle$ in the BL region might arise from locally varying production rates. Indeed, $\overline{Nu}$, and hence also the temperature gradient within the BL, is highest for $d<0$, which coincides with the strongest amplitude in $\langle \theta_S^2\rangle$. Statistics of $\theta$ and $w$ are interconnected since temperature fluctuations drive production of $\langle w^2\rangle$. Wall-normal velocity fluctuations then feed back to  temperature fluctuations since the term $\theta w$ also features in the production term for $\langle \theta_S^2\rangle$. This picture is consistent at least with our results for $\rho_\theta$, $\rho_w$ as well as for $\rho_{\theta w}$ in the near wall-region. To what extent this theory actually applies will have to be evaluated in future studies. One way to address this would be to study the $Ra$ dependence of the effects described here. Recent studies in 2D \citep{Zhu2018} and 3D RBC \citep{Blass2020} indicate that the $\overline{Nu}$ distribution changes with increasing $Ra$. This should lead to changes in the small-scale patterns if the above is true. A complete answer is likely even more complicated since the different scalings identified in \citet{He2019} are linked to different leading order balances (inertia-viscous \emph{vs.} inertia-bouyancy) \citep{Adrian1996}. Naturally, such a fundamental difference is also to have an impact on the quantities studied here. 

A further relevant question is how the strength of the scale interaction varies as a function of $Ra$. Experience from flat plate turbulent boundary layers shows that in this case the modulation of small scales by the large superstructure scales intensifies with increasing shear Reynolds number $Re_\tau$ \citep[e.g.][]{mat09}. It remains to be seen if a similar trend exists in RBC. This will also shed light on whether a `universal small-scale signal' exists in pure convection. That is, if the properly normalized statistics of the small-scale signal at a given $z$ are invariant for different locations within the LSC and over at least a limited range of $Ra$ once they are stripped of the large-scale footprint by `demodulation'. Identifying such a universal signal will be crucial in representing small-scale turbulence in RBC effectively.

\rev{Finally, it would be interesting to uncover the $Pr$-dependence of the effects discussed here. Changes in $Pr$ affect the small scale signal as well as the LSC, rendering it difficult to make any predictions on their interaction based on the present results. }

\section*{Acknowledgements}
We are grateful to Richard Stevens for making the data available to us and thank Roberto Verzicco and Detlef Lohse for valuable feedback and support. This project is supported by the Priority Programme SPP 1881 Turbulent Superstructures of the Deutsche Forschungsgemeinschaft and by the University  of  Twente  Max-Planck  Center  for  Complex  Fluid Dynamics. Part of the work
was carried out on the national e-infrastructure of SURFsara, a subsidiary of SURF
cooperation, the collaborative ICT organization for Dutch education and research.

\section*{Declaration of Interests}
The authors report no conflict of interest.

\section*{\rev{Appendix A: Details of the wavelet transform used in \S\,\ref{ssec:red}}}
\rev{In decomposing the 2D spatial field $\xi(x,y)$ in space-scale space, the 2D continuous wavelet transform (CWT) algorithm, \texttt{cwtft2}, of the Matlab$^\copyright$ Wavelet Toolbox was utilized. This algorithm efficiently performs the convolution (\ref{eq:wavconv}) in the wavenumber domain, and is repeated for a range of spatial wavelet scales $l_s$. Following Plancherel's theorem the equivalent of the operation defined in (\ref{eq:wavconv}) in wave number space is given by (for an isotropic wavelet, thus without angular dependency):
\begin{equation}
A_{\xi}(x,y;l_s) = \int\int \widehat{\xi}(k_x,k_y)l_s\widehat{\overline{\Psi}}\left(l_sk_x,l_sk_y\right){\rm d}k_x{\rm d}k_y,
\end{equation}
where the hat signifies the Fourier transform, and $k_x$ and $k_y$ are the horizontal wavenumbers, which are related to the radial wavenumber by $k \equiv \sqrt{k_x^2 + k_y^2}$. An isotropic and admissible Morlet wavelet was chosen, defined in wavenumber space as
\begin{equation}
\widehat{\Psi}\left(k_x,k_y\right) = e^{-0.5\left(\vert k_x+i k_y\vert - k_0\right)^2},
\end{equation}
with $k_0 = 6$. Alternative choices of isotropic wavelets (such as the Mexican hat wavelet) with different resolution characteristics in wavelet scale \emph{vs.} physical space were found to not affect the conclusions made throughout the manuscript. This is related to the fact that the wavelet transform-results are used in terms of the (large-scale) variation in the small-scale energy, being $E_\xi(x,y;k_s > k_\mathrm{cut})$. Here, the scale separation between the large- and small-scales is large (recall figure~\ref{fig:spec} and its discussion) and is thus unaffected by a wavelet transform resulting in more or less resolution in wavelet scale or physical space, respectively (or vice versa).}

\bibliographystyle{jfm}
\bibliography{amp_mod_jfm}

\end{document}